\documentclass[showpacs,twocolumn,prd,superscriptaddress]{revtex4-1}

\usepackage{slashed}
\usepackage{mathrsfs}
\usepackage{amsmath}
\usepackage{amssymb}
\usepackage{revsymb}
\usepackage{graphicx,accents}
\usepackage{graphicx,epstopdf}
\usepackage{mathrsfs}
\usepackage{bm}
\usepackage{psfrag}
\usepackage{color}
\usepackage{hyperref}
\hypersetup{colorlinks=true,citecolor=Red,urlcolor=Red}
\usepackage{bbm}
\usepackage{bbold}
\usepackage{tabularx}
\usepackage{graphicx}
\usepackage[normalem]{ulem}
\usepackage[usenames,dvipsnames]{xcolor}

\usepackage[usenames,dvipsnames]{xcolor}
\usepackage{multirow,tabularx}
\usepackage{xcolor,pifont}
\newcommand*\colourcheck[1]{%
  \expandafter\newcommand\csname #1check\endcsname{\textcolor{#1}{\ding{52}}}%
}

\newcommand{\be}{\begin{equation}}
\newcommand{\ee}{\end{equation}}
\newcommand{\bea}{\begin{eqnarray}}
\newcommand{\eea}{\end{eqnarray}}
\newcommand{\ud}{\mathrm{d}}
\newcommand{\eps}{\varepsilon}
\newcommand{\mbf}[1]{\mathbf{#1}}
\newcommand{\trm}[1]{\textrm{#1}}

\newcommand{\figref}[1]{Fig. \ref{#1}}

\newcommand{\eqnref}[1]{Eq.\,(\ref{#1})}

\newcommand{\tsf}[1]{\textsf{#1}}

\newcommand{\vkap}{\varkappa}
\newcommand{\vphi}{\varphi}
\newcommand{\vtheta}{\vartheta}

\newcommand{\Ai}{\trm{Ai}}

\newcommand{\nn}{\nonumber}

\newcommand{\LCperp}{{\scriptscriptstyle \perp}}

\newcommand{\e}{\mbox{e}\,}

\newcommand{\sscript}{\scriptscriptstyle}

\begin{document}

%%%%%%%%%%%%%%%%%%%%%%%%%%%%%%%%%%%%%%%%%%%%%%%%%%

\title{Nonlinear Compton scattering of polarised photons in plane-wave backgrounds}

\author{B.~King}
\email{b.king@plymouth.ac.uk}
\affiliation{Centre for Mathematical Sciences, University of Plymouth, Plymouth, PL4 8AA, United
Kingdom}
\author{S.~Tang}
\email{suo.tang@plymouth.ac.uk}
\affiliation{Centre for Mathematical Sciences, University of Plymouth, Plymouth, PL4 8AA, United
Kingdom}

%%%%%% PACS:

\date{\today}
\begin{abstract}
We investigate the phenomenology of nonlinear Compton scattering of polarised photons by unpolarised electrons in plane-wave backgrounds. The energy and angular spectra of polarised photons are calculated for linearly- and circularly-polarised pulses, monochromatic fields and constant crossed field backgrounds. When the field intensity is in the weakly nonlinear regime, photons in different polarisation states are predicted to possess very different energy and angular distributions. We explain this difference
by calculating the spectrum for nonlinear Thomson scattering in a linearly-polarised monochromatic background and projecting the electron's trajectory onto the different polarisation directions. Finally, we calculate a scenario for multi-GeV electrons and by considering the energy and angular-resolved photon polarisation purity, find one can achieve a GeV-photon beam with polarisation purities of over $90\%$.
\end{abstract}
\maketitle
\twocolumngrid

\section{Introduction}

The process in which an electron collides with a laser pulse and absorbs more than a single laser photon before radiating is referred to as nonlinear Compton scattering (NLC): $e \to e + \gamma$. It has been studied theoretically in a constant crossed background~\cite{nikishov64}, a monochromatic plane wave~\cite{nikishov64,kibble64,serbo04,harvey09}, a finite pulse~\cite{boca09,mackenroth10,heinzl10b,mackenroth11,seipt11} and adapted to numerical Monte Carlo simulations \cite{elkina11,king13a,ridgers14,harvey15b,gonoskov15}. Experimentally, up to the fourth harmonic has been observed in the E144 experiment \cite{bula96}, which collided a high-energy electron of $46.6\,\trm{GeV}$ with an intense laser pulse \cite{bamber99}. Evidence of the quantum effect of recoil in radiation reaction has recently been seen in laser-plasma experiments \cite{cole18,poder18} and experimental campaigns in E320 at FACET-II \cite{slacref1} and in LUXE at DESY \cite{Abramowicz:2019gvx} plan to measure the transition of NLC to the non-perturbative regime.

There has been recent attention given to the description of electron and photon polarisation in NLC. Using a non-precessing polarisation basis \cite{king15b}, one can calculate the rate of polarisation flip in single NLC events, which have been added to laser-plasma simulation codes \cite{delsorbo17}. It is necessary to introduce some asymmetry in the background field to prevent the spin-polarising effect of the laser pulse from averaging out to zero and various schemes such as counterpropagating pulses \cite{delsorbo18} or using an elliptically-polarised beam \cite{yanfei19} have been suggested to achieve higher spin-polarisation purities.
Using a simulation approach based on the locally constant field approximation (LCFA) \cite{ritus85,king15d,dipiazza18,dipiazza19,king19a,king19d} in the quantum-dominated radiation-reaction regime \cite{dipiazza10}, it was recently suggested \cite{li19} that using high energy \emph{spin-polarised} electron beams colliding with an intense laser pulse, and an energy cut to the emitted photons, highly polarised photon beam could be achieved. At the same time as the current manuscript was in preparation, \cite{wistisen19} performed a QED calculation valid for arbitrary pulse length and spin-polarised electrons in a circularly-polarised background, which concluded that the deterioration in photon polarisation purity that accompanies increasing nonlinearity parameter can be compensated for by also using an angular cut.

A further importance of polarised single (dressed) vertex processes is their occurrence in higher-order processes. It has been shown several times \cite{ritus72,baier72,king13b,king18c,torgrimsson18} that to correctly factorise the ``two-step'' trident process of NLC followed by nonlinear Breit-Wheeler pair-creation, observed in the E144 experiment \cite{burke97,hu10}, one is required to consider the transverse polarisation of the intermediate photon (whereas the ``one-step'' process involves all polarisation states of the photon \cite{ilderton11,mackenroth18}). Analogously, in double NLC, it is the polarisation of the intermediate electron that must be calculated for a correct factorisation \cite{morozov75,seipt12,king15b}, and this ``gluing'' approach has recently been extended to cascades \cite{torgrimsson19}. Therefore it is of use to have expressions for polarised processes in a range of backgrounds, which also can be employed in higher-vertex calculations.

In the current paper, we focus on the polarisation of the emitted photons in NLC and focus on the high-energy, intermediate intensity regime that is planned to be probed in E320 \cite{slacref1} and LUXE \cite{Abramowicz:2019gvx}. Being able to prepare high-energy photon beams with a high degree of polarisation purity is advantageous for phenomenological studies. For example, it has been suggested that combining polarised high-energy photons with currently-available laser facilities would allow the first measurement of real photon-photon scattering, either directly \cite{ilderton16,king16,homma15,meuren17b}, or via the Kramer-Kronig relation that links the process to polarised Breit-Wheeler pair-creation \cite{Toll:1952rq,wistisen20}. A general framework for incorporating polarised NLC to numerical simulations was derived in~\cite{king13a} and applied to study electromagnetic cascades in rotating electric fields.

We derive photon-polarised NLC from unpolarised electrons in a pulsed plane-wave background, and show that the energy and angular spectra of the emitted photons depend sensitively on their polarisation states. Therefore, by using angular and energy cuts, one can achieve a higher degree of polarisation than has so far been reported in the literature.
We find that the angular distribution of the photons in different polarisation states are qualitatively different, exhibiting typical patterns of the different orders of multipole radiation. By deriving photon-polarised NLC in circularly- and linearly-polarised monochromatic backgrounds and comparing with a classical analysis of the electron trajectory, we find an explanation for the familiar multipole structures in the angular spectrum that can be traced to the projection of the electron's motion on the photon polarisation directions.

The paper is organised as follows.
Sec.~\ref{Sec_2} gives an overview of the finite pulse derivation and defines important quantities that will be used throughout the paper.
In Sec.~\ref{Sec_linear}, we outline the derivations for the production of linearly polarised photons in linearly polarised backgrounds: a pulsed field, a monochromatic field and a constant crossed field and then in Sec.~\ref{Sec_circular}, we proceed with the production of circularly-polarised photons in circularly polarised backgrounds: a pulsed field, a monochromatic field and a constant crossed field.
In Sec.~\ref{Sec_numerical}, we display numerical results that show the significant discrepancy in the polarised photon energy spectrum and angular distribution by the QED result compared with the LCFA.
In Sec.~\ref{Sec_classical} we outline a classical calculation of the equivalent number of photons in the radiated field, and demonstrate how, by decomposing the classical trajectory along the photon polarisation directions, we recover the leading order behaviour of the QED result. In Sec.~\ref{Sec_conclusion} we conclude. Appendix~\ref{AppendixA} contains a derivation of the regularisation used for some of the photon-polarisation-dependent terms arising from the trace.

\section{Scattering in pulse background}~\label{Sec_2}
All the electromagnetic (EM) backgrounds we consider will be of transverse plane-wave nature with a reduced vector potential $a(\phi)= eA(\phi)$, for positron charge $e$ and vector potential $A$, and $\phi=\vkap\cdot x$ is the external-field phase with wavevector $\vkap=\omega_{0}(1,0,0,1)$ satisfying $\vkap^{2}=0$, $\vkap\cdot a = 0$, and $\omega_0$ is the field frequency. The scattering-matrix element for NLC is then:
\bea
\tsf{S}_{\tsf{fi}} =-ie~\int \ud^{4}x~\overline{\psi}_{q} \slashed{\epsilon}^{\ast} \psi_{p}\frac{\e^{i k\cdot x}}{\sqrt{2Vk^{0}}},
\eea
where $p$ and $q$ are the incoming and outgoing electron momenta respectively, $k$ and $\epsilon$ are the scattered photon momentum and polarisation respectively, satisfying $k\cdot \epsilon = 0$, and we recap the Volkov state
\bea
\psi_{p} &=& \left[1+ \frac{\slashed{\vkap}\slashed{a}(\phi)}{2\,\vkap\cdot p}\right]\,\frac{u_{p}}{\sqrt{2p^{0}V}} \e^{-ip\cdot x - i\int^{\phi} \frac{2 p\cdot a(\vphi) - a^{2}(\vphi)}{2\,\vkap\cdot p}~\ud\vphi}. \nn
\eea

We wish to express the photon polarisation in terms of the eigenstates of the polarisation operator in the background field, which ensures photon polarisation will not change after the emission. (The probability for photon-photon scattering \cite{king15a} is very small, so that deviations in experiment from the plane-wave background are not expected to appreciably impact our assumption that the photon polarisation remains constant after emission.) In linear background fields with polarisation $\eps_{1}=(0,1,0,0)$ [or $\eps_2=(0,0,1,0)$], the eigen-polarisation can be expressed in terms of the orthogonal basis \cite{baier75a}:
\begin{align}
\epsilon_{1} = \eps_{1} - \frac{k\cdot \eps_1}{k\cdot \vkap} \vkap\,,~~~&\epsilon_{2} = \eps_2 - \frac{k\cdot \eps_2}{k\cdot \vkap} \vkap\,,\nonumber\\
\mbox{e}_{\pm} =  \pm k &+ \frac{1}{2\,k\cdot \vkap} \vkap\,,
\label{eqn:polbasis}
\end{align}
which satisfies the normalisation: \mbox{$\epsilon_{1}^{2}=\epsilon_{2}^{2}=\mbox{e}_{-}^{2}=-\mbox{e}_{+}^{2}=-1$}, and up to a sign, the photon transverse polarisation vectors are related to the invariants
\[
\epsilon_{1} \sim \frac{k \cdot F}{\sqrt{-(k\cdot F)^{2}}}; \quad \epsilon_{2} \sim \frac{k\cdot \widetilde{F}}{\sqrt{-(k\cdot \widetilde{F})^{2}}},
\]
where $F$ is the Faraday field tensor and $\widetilde{F}$ its dual.
Whereas in circular background fields with polarisation $(\eps_{1} \pm i\eps_{2})/\sqrt{2}$, the eigenbases $\epsilon_{1,2}$ are replaced with
\[\epsilon_{1,2}\to \epsilon_{\pm} = \frac{1}{\sqrt{2}}(\epsilon_{1}\pm i \epsilon_{2})\,.\]

We then calculate the probability $\tsf{P}_{j}$ for NLC of a photon with polarisation $\epsilon_{j}$ to be:
\bea
\tsf{P}_{j} = V^{2} \int \langle |\tsf{S}_{\tsf{fi}}(\epsilon_{j})|^{2}\rangle_{\trm{spin}}~ \frac{\ud^{3}q\,\ud^{3}k}{(2\pi)^{6}},
\eea
where $\langle\cdot\rangle_{\trm{spin}}$ refers to averaging over initial and summing over final spins of the electron, $j=1,2$ in the linear background field and $j=+,-$ in the circular field.
Expanding the photon polarisation in the basis \eqnref{eqn:polbasis}, only the photon's transverse polarisation states ($\epsilon_{1}$ and $\epsilon_{2}$ in linear backgrounds, $\epsilon_{+}$ and $\epsilon_{-}$ in circular backgrounds) will survive, allowing us to write the total probability $\tsf{P}$ as a sum of the probability of scattering into each of these polarisation states $\tsf{P} = \tsf{P}_{1}+\tsf{P}_{2}$ ($\tsf{P} = \tsf{P}_{+}+\tsf{P}_{-}$) for linear (circular) background fields.
Since the only non-trivial dependency on $x$ is in the dependency on $\vkap \cdot x$, one acquires a momentum-conserving delta function in three co-ordinates, and we choose to integrate out the scattered electron momentum $q$. Eventually we arrive at the intermediate stage:
\bea
\tsf{P}_{j} &=& \frac{\alpha}{(2\pi)^{2}}\frac{1}{\eta_{p}^{2}} \int_{0}^{1} \frac{s}{t}\ud s\int \ud^{2}r^{\LCperp} \int\ud\phi\,\ud\phi'\nn \\
&&\times \left[\tsf{T}(\epsilon_j)\e^{i c(\phi-\phi') -i\int^{\phi}_{\phi'} \frac{2 p\cdot a(\vphi) - a^{2}(\vphi)}{2\,\vkap\cdot p}~\ud\vphi} \right], \label{eqn:sfi9}
\eea
where $\alpha=e^2/4\pi$ is the fine structure constant, \mbox{$\eta_{p} = \vkap \cdot p / m^{2}$},
$s=\vkap\cdot k / \vkap \cdot p$ is the lightfront momentum fraction of the scattered photon, $t=1-s$ and $c = (k^{+}+q^{+}-p^{+})/2\vkap^{0}$. We use lightfront co-ordinates:
$x^{\pm} = x^{0} \pm x^{3},~x^{\LCperp} = (x^{1}, x^{2})$ and $r^{\LCperp}=k^{\LCperp}/sm$ is the normalised photon momentum in the plane perpendicular to the laser propagation direction, and relates directly to the outgoing angle of the scattered photon:
\begin{align}
r^{\LCperp}=\frac{m\eta_p}{\omega_{0}}\frac{\sin\theta}{1+\cos\theta}(\cos\psi,\sin\psi)\,,
\label{Eq_scattering_angle}
\end{align}
where $\theta$ is the polar angle measured from a head-on collision of the photon with the laser pulse, and $\psi$ is the azimuthal angle. The dependence of the emission probability $\tsf{P}_j$ on the photon polarisation state $\epsilon_j$ is included in:
\[
\tsf{T}(\epsilon_j) =  \overline{E}_{q}(\phi) \slashed{\epsilon}^{\ast}_{j} E_{p}(\phi)\overline{E}_{p}(\phi')\slashed{\epsilon}_{j}E_{q}(\phi'),
\]
and we use the shorthand:
\[
E_{p}(\phi) = \left[1+ \frac{\slashed{\vkap}\slashed{a}(\phi)}{2\,\vkap\cdot p}\right]\,\frac{u_{p}}{\sqrt{2p^{0}V}}.
\]

\section{Linearly Polarised Photon}~\label{Sec_linear}
A linearly-polarised background field has linearly-polarised photon eigenstates. Here, we consider a laser background of the form:
\bea a(\phi)=m\xi g(\phi)\sin(\phi)\eps_{1}\,, \label{eqn:alinear1}\eea
which has polarisation $\eps_1$, intensity parameter $\xi$ and pulse envelope $g(\phi)$.

The derivation is very similar to the well-known unpolarised case~\cite{nikishov64,kibble64}. Performing the calculation for $\tsf{T}(\epsilon_{j})$, we arrive at the linearly polarised photon-momentum spectrum:
\begin{align}
\frac{\ud^3\tsf{P}_{j}}{\ud s \ud^2r^{\LCperp}}=&\frac{\alpha}{(2\pi)^2} \frac{1}{\eta^2_p} \frac{s}{t}  \int \ud\phi \int \ud\phi'e^{i(\phi-\phi')\big\langle\frac{k\cdot \pi_p}{\vkap\cdot q}\big\rangle}\nonumber\\
\times&\left[\Pi(\phi)\cdot\varepsilon_{j}~\Pi(\phi')\cdot\varepsilon_{j}+\Delta\frac{s^2}{8t}\right]\,,
\label{eqn:pshortang1}
\end{align}
where we define the instantaneous classical electron momentum in a plane wave:
\[\pi_{p}(\phi) = p - a(\phi) + \vkap\,\frac{2p\cdot a(\phi) - a^{2}(\phi)}{2\vkap \cdot p}\,,\]
and use the shorthand: $\Delta = [a(\phi)-a(\phi')]^2/m^{2}$, \mbox{$\Pi(\phi) = \pi_p(\phi)/m - k/ms$}.
The window average of $f$ is denoted by $\langle f \rangle = (\phi-\phi')^{-1}\int_{\phi'}^{\phi}f(\vphi)\ud\vphi$. \eqnref{eqn:pshortang1} contains a (divergent) contribution from a pure phase term, which persists outside the laser pulse, and therefore must be regularised. Since we are interested in angular distributions, we cannot use the standard ``$i\epsilon$'' prescription~\cite{dinu14a} but instead must regularise in the way introduced in \cite{boca09} and recently further developed in \cite{king19d}. This then leads to inserting regulation factors $R(\phi)$ [$R(\phi')$]:
\bea
R(\phi) = 1 - \frac{k\cdot \pi_{p}(\phi)}{k\cdot p}\,,
\label{Eq_regular_factor}
\eea
wherever necessary in Eq.~(\ref{eqn:pshortang1}) to ensure the integrand is zero outside the pulse, which means $\Pi(\phi) \to \Pi_{\trm{reg.}}(\phi)$:
\bea
\Pi_{\trm{reg.}}(\phi) = -\ell R(\phi) - a(\phi)/m,
\eea
and $\ell$ is the normalized shifted momentum:
\bea
\ell = \frac{k}{sm} - \frac{p}{m}\,. \label{eqn:rdef}
\eea

So far this derivation has been quite general, we now specify it in a standard way to short pulses by defining an average $\vphi=(\phi+\phi')/2$ and interference phase $\vartheta=\phi-\phi'$~\cite{king19a}. Integrating over the transverse momentum $k^{\LCperp}$, we arrive at the photon-polarised lightfront-momentum spectrum: \mbox{$\tsf{P}_{1,2} = (\tsf{P}\pm \Delta\tsf{P})/2$}, where $\tsf{P} = \tsf{P}_{1} + \tsf{P}_{2}$ is the total unpolarised probability and $\Delta\tsf{P}=\tsf{P}_{1}-\tsf{P}_{2}$ is the change in probability due to scattering into a particular polarisation state:
\begin{subequations}
\begin{align}
\frac{\ud\tsf{P}}{\ud s} &=\frac{\alpha}{\pi\eta_{p}} \int \ud\vphi\,\left(\int^{\infty}_{0} \frac{\ud\vtheta}{\vtheta}~\mathcal{K}-\frac{\pi}{2}\right),\\
\frac{\ud\Delta\tsf{P}}{\ud s} &= \frac{\alpha}{\pi\eta_{p}} \int \ud\vphi\int^{\infty}_{0} \frac{\ud\vtheta}{\vtheta} \left(\mathfrak{a}_{1}-\mathfrak{a}_{2}\right)\sin \left(\frac{ s\vtheta\mu}{2\eta_{p} t}\right),
\end{align}
\label{Eq_lin_Perp_integral}
\end{subequations}
where
\[\mathcal{K} = \left(1 - \Delta\frac{1+t^{2}}{4t}\right) \sin \left(\frac{s\vtheta\mu}{2\eta_{p} t}\right)\,,\]
the Kibble mass \mbox{$\mu= \mu(\vphi,\vtheta)= 1 - \langle a^{2} \rangle /m^{2}+ \langle a \rangle^{2}/m^2$},
and $\mathfrak{a}_{j} = \left[\langle a \cdot \eps_{j} \rangle - a(\phi) \cdot \eps_{j}\right] \left[\langle a \cdot \eps_{j} \rangle - a(\phi') \cdot \eps_{j}\right]/m^2$. One significant difference in the standard derivation of NLC in a pulse, is that, when one considers polarised electrons or photons, a term arises of the form:
\bea
\int \frac{\ud\vtheta}{\vtheta^{2}}\,\mbox{e}^{if(\vtheta)}
\eea
in addition to the usual terms with pre-exponent integrands of the form $1/\vtheta$. The regularisation of this term is given in Appendix \ref{AppendixA}.

\subsection{Photon-polarised NLC LCFA}
The locally constant field approximation (LCFA) can be acquired from Eq.~(\ref{eqn:pshortang1}) by performing a Taylor series in the interference phase. To make the connection, let us specify the \emph{field} to be $a'(\vphi) = m\xi(\vphi)\,\eps_1$, where $\xi(\vphi)$ includes the pulse amplitude and waveform. Then making the substitutions:
\begin{align}
\Delta  &\to -\vtheta^{2}\xi^{2}(\vphi)\,,\nn\\
\vtheta\left\langle\frac{k\cdot \pi_p}{\vkap\cdot q}\right\rangle &\to \frac{k\cdot \pi_p(\vphi)}{t\eta_p}\vtheta + \frac{1}{24}\frac{ s\xi^2(\vphi)}{t\eta_p}\vtheta^3\,,\nn\\
\Pi(\phi)\cdot\varepsilon_{j}\,\Pi(\phi')\cdot\varepsilon_{j} &\to \left[\Pi(\vphi)\cdot\varepsilon_{j}\right]^2-\vtheta^2\frac{\xi^2(\vphi)}{4} \delta_{1,j}\,,\nn
\end{align}
in Eq.~(\ref{eqn:pshortang1}), one can simply perform the integral over $\vtheta$ to acquire the polarised photon triple differential spectrum:
\begin{align}
\frac{\ud^{3} \tsf{P}_{j}}{\ud s\,\ud^2r^{\LCperp}}=&\frac{\alpha}{\pi \eta_p}\int \ud\vphi~\textrm{Ai}(y)\left[\left[\Pi(\vphi)\cdot\varepsilon_{j}\right]^2z+\delta_{1,j}y +\frac{s^2}{t}y\right] \nn\\
\label{Eq_NCS_Lin_LCFA}
\end{align}
where
\[y=\frac{2}{s}\frac{k\cdot \pi_{p}(\vphi)}{m^2}z, \qquad z=\left(\frac{s}{\chi_{p}t}\right)^{2/3} \]
and $\chi_{p}=\chi_{p}(\vphi) = |\xi(\vphi)|\eta_{p}$.
We note that this angular-resolved LCFA result depends not only on the local electric field $\xi(\vphi)$, but also on the local vector potential $a(\vphi)$ included in $\pi(\vphi)$. Integrating over transverse momenta $k^{\LCperp}$, the lightfront momentum spectrum becomes:
\begin{subequations}
\bea
\frac{\ud\tsf{P}}{\ud s} &=&-\frac{\alpha}{\eta_{p}}\int \ud\vphi  \left[\Ai_{1}(z) + \frac{1+t^2}{t}\frac{1}{z} \Ai'(z)\right], \\
\frac{\ud\Delta\tsf{P}}{\ud s} &=&~~\frac{\alpha}{\eta_{p}}\int \ud\vphi  \,\frac{1}{z}\Ai'(z)\,,
\eea
\label{Eq_lin_LCFA_S}
\end{subequations}
which is just an integration over the pulse shape of the NLC result in a constant crossed field \cite{king15b}.

\subsection{Linearly-polarised monochromatic background}
Beginning with \eqnref{eqn:pshortang1}, we derive photon-polarised NLC for a linearly polarised monochromatic background [corresponding to $g(\phi) = 1$ in \eqnref{eqn:alinear1}]. The main complication introduced by linear polarisation of the background is that $a^{2}$ is not constant. This means that one is faced with squared sinusoidal function in the exponent and pre-exponent. These can be handled straightforwardly by a Fourier decomposition. To this end, let us define the functions $\Gamma_{l,n}$ via:
\bea
 \cos^{h}\phi~\mbox{e}^{i\left(-\zeta_{\sscript{l}}\sin\phi + \beta \,\sin2\phi\right)} = \sum_{n=-\infty}^{\infty} \Gamma_{h,n}\mbox{e}^{-in\phi}.
\eea
where
\bea
\zeta_{\sscript{l}}=\frac{s\xi}{t\eta_p}\ell\cdot \varepsilon_{1}\,,~~~\beta = \frac{s\xi^2}{8t\eta_p}\,.
\eea
We then find that $\Gamma_{h,n}=\sum_{m=-\infty}^{\infty}J_{m}(\beta)F_{h,m,n}(\zeta_{\sscript{l}})$ with
\bea
 F_{0,m,n}&=&J_{2m+n}(\zeta_{\sscript{l}})\,, \nn \\
 F_{1,m,n}&=&\frac{1}{2}\left[\,J_{2m+n+1}(\zeta_{\sscript{l}})+J_{2m+n-1}(\zeta_{\sscript{l}})\right]\,,\nn\\
 F_{2,m,n}&=&\frac{1}{4}\left[\,J_{2m+n+2}(\zeta_{\sscript{l}})+2\,J_{2m+n}(\zeta_{\sscript{l}})+J_{2m+n-2}(\zeta_{\sscript{l}})\right]. \nn
\eea
where $J_{n}(\zeta_{\sscript{l}})$ is the Bessel function.
Unlike for the pulsed background case, there is no need to introduce an explicit regularisation (being infinite in extent, there is no place where the monochromatic background disappears). Finally, we find $\tsf{P}_{j} = \sum_{n=-\infty}^{\infty} \tsf{P}_{j,n}$, where
\begin{align}
&\tsf{P}_{j,n}=\frac{\alpha N_{\phi}}{\eta_p^2} \int^{1}_{0}\ud s\int \ud^2r^{\LCperp}~\frac{s}{t}\,\delta(c+2\beta-n)\label{eqn:Pjn2}\\
     &\times\left[\left(\ell\cdot \varepsilon_{j}\,\Gamma_{0,n}-\delta_{1j}\xi\, \Gamma_{1,n}\right)^2-\left( \Gamma_{2,n} \Gamma_{0,n}-\Gamma^2_{1,n}\right) \frac{s^2\xi^2}{4t} \right],\nonumber 
\end{align}
and we label $N_{\phi} = \delta(x)|_{x\to 0} \equiv \int d\phi /2\pi$ as the number of laser cycles, which is formally infinite, but which we will subsequently set equal to a finite value for the purpose of comparison with the pulsed result. The delta-function:
\bea
 \delta(c+2\beta-n) = \delta\left[\frac{s}{2\eta_{p}t} \left(1+ \frac{\xi^{2}}{2}+ \ell^2_{\LCperp}\right) - n \right]~~ \label{eqn:delt1}
\eea
fixes $n>0$ (where $n$ is often interpreted as the net number of absorbed laser photons) and also gives the kinematic range of $n$th harmonic~\cite{harvey09}, $0<s<s_{{\sscript l},n}$, where:
\bea
s_{{\sscript l},n}=\frac{2n\eta_{p}}{2n\eta_{p}+1+ \xi^{2}/2}\,.
\label{Eq_harmonic_region_lin}
\eea

\subsubsection{Linearly-polarised angular spectrum}
Integrating out $s$ in Eq.~\ref{eqn:Pjn2} gives the angular dependency:
\begin{align}
\tsf{P}_{j,n}=&\frac{\alpha N_{\phi}}{n\eta_p^2}\int\frac{\ud^2r^{\LCperp}\,\mathcal{L}^2_{n} }{(1+\mathcal{L}_n)^2}\left[\left(\ell\cdot \bm{\varepsilon}_{j}\Gamma_{0,n}-\delta_{1j}\xi\Gamma_{1,n}\right)^2\right.\nonumber\\
 &\left.~~~-\left(\Gamma_{2,n} \Gamma_{0,n}-\Gamma^2_{1,n}\right) \frac{\mathcal{L}^2_n\xi^2}{4(1+\mathcal{L}_n)} \right]\,,
\end{align}
 where
\bea
\mathcal{L}_{n} = \frac{2n\eta_{p}}{(\ell^{\LCperp})^{2}+1+\xi^{2}/2}\label{eqn:yast1}
\eea
and the arguments of the $\Gamma$-functions have become:
 \[
 \zeta_{{\sscript l}} \to \zeta_{{\sscript l},n}= \frac{\xi\,\mathcal{L}_{n}}{\eta_{p}}\ell\cdot \varepsilon_1; \qquad \beta\to\beta_n = \frac{\xi^{2}\,\mathcal{L}_{n}}{8\eta_{p}}.
 \]

\subsubsection{Linearly-polarised lightfront momentum spectrum}
If \eqnref{eqn:delt1} is used to integrate out the perpendicular photon momentum instead, we acquire:
\bea
\tsf{P}_{j,n}&=&\frac{\alpha N_{\phi}}{\eta_p} \int^{s_{{\sscript l},n}}_{0}\ud s\int^{\pi}_{-\pi} \ud\theta_{r}\left[\left(\ell_{n}\cdot \varepsilon_{j} \Gamma_{0,n}-\delta_{j,1}\xi\Gamma_{1,n}\right)^2\right.\nonumber\\
 &&\qquad\qquad\qquad\left.-\left(\Gamma_{2,n} \Gamma_{0,n}-\Gamma^2_{1,n}\right) \frac{s^2\xi^2}{4t} \right]\,,
 \label{eqn:quant1}
\eea
where $\ell_n\cdot\eps_{1} =-|\ell^{\LCperp}_{n}|\cos\theta_{r}$, $\ell_n\cdot\eps_{2} =-|\ell^{\LCperp}_{n}|\sin\theta_{r}$,
and we define $|\ell_{n}^{\LCperp}| = \sqrt{2n\eta_{p} (s_{{\sscript l},n}-s)/(s\,s_{{\sscript l},n})}$. The argument $\zeta_{{\sscript l}}$ of the $\Gamma$-functions becomes:
 \[
 \zeta_{{\sscript l}} \to \zeta_{{\sscript l},n}=-\frac{s\,\xi}{t\eta_{p}}|\ell^{\LCperp}_{n}|\cos\theta_{r}.
 \]

\section{Circularly polarised photon}~\label{Sec_circular}
In this section, we consider the background to be a circularly polarised field:
\begin{align}
a(\phi)&=m\xi g(\phi)[\varepsilon_1\cos(\phi)+\varepsilon_2\sin(\phi)],\label{eqn:clinear1}
\end{align}
which has a right-handed polarisation:
\[
\eps_{+} = \frac{1}{\sqrt{2}}\left(\eps_{1} + i \eps_{2}\right)\,.
\]
The derivation is very similar to the linearly-polarised case. Repeating the previous derivation and replacing the photon polarisation with $\epsilon_{\pm}$, we acquire the circularly polarised photon-momentum spectrum:
\begin{align}
&\frac{\ud^{3}\tsf{P}_{\pm}}{\ud s \ud^2r^{\LCperp}}=\frac{\alpha}{(2\pi\eta_p)^2} \frac{s}{t} \int \ud\phi \int\ud\phi' e^{i(\phi-\phi')\big\langle\frac{k\cdot \pi_p}{\vkap\cdot q}\big\rangle}\times\nonumber\\
  &~\frac{1}{2}\left[\frac{s^2}{4t}\Delta+\Pi^{\LCperp}(\phi)\cdot\Pi^{\LCperp}(\phi')\pm if_{s}\Pi^{\LCperp}(\phi)\times \Pi^{\LCperp}(\phi')\right],
\label{Eq_NCL_cir_probability}
\end{align}
in which we define $f_{s}=1+s^2/2t$, and use the shorthand: \mbox{$x^{\LCperp}\cdot y^{\LCperp}=x^{1}y^{1}+x^{2}y^{2}$} and \mbox{$x^{\LCperp}\times y^{\LCperp}=x^{1}y^{2}-x^{2}y^{1}$}. We employ here the same regularisation as in the linear case: $\Pi^{\LCperp}(\phi)\to \Pi^{\LCperp}_{\trm{reg.}}(\phi)$. Performing the integration over the transverse momentum, we acquire the light-front momentum spectrum:
\begin{align}
\tsf{P}_{\pm}=&\frac{\alpha}{2\pi \eta_p}\int^{1}_{0} ds\int \ud\vphi \left(\int^{\infty}_{0} \frac{\ud\theta }{\theta}~\mathcal{K}_{\pm}-\frac{\pi}{2}\right)\,,
\label{Eq_cir_Perp_integral}
\end{align}
where
\begin{align}
\mathcal{K}_{\pm}=\left[1-\frac{1}{2}f_{s} \Delta \pm i f_{s}\left(\mathfrak{b}-\mathfrak{b}^{*}\right)\right]\sin\left(\frac{s\theta\mu}{2t\eta_p}\right)\,,
\end{align}
where \mbox{$\mathfrak{b}=[a(\phi')-\langle a\rangle]\cdot\varepsilon_{1}~[a(\phi)-\langle a\rangle]\cdot\varepsilon_{2}/m^2$} and \mbox{$\mathfrak{b}^{*}=[a(\phi)-\langle a\rangle]\cdot\varepsilon_{1}~[a(\phi')-\langle a\rangle]\cdot\varepsilon_{2}/m^2$}. Summing over polarisations in Eq.~(\ref{Eq_cir_Perp_integral}) then recovers the same polarisation-averaged probability, $\tsf{P}=\tsf{P}_{+}+\tsf{P}_{-}$, as in the linear polarisation case, Eq.~(\ref{Eq_lin_Perp_integral}).

\subsection{Photon-polarised NLC LCFA}
With the same procedure as in the linear case, we acquire the LCFA result for circular polarisation:
\begin{align}
&\frac{\ud^3\tsf{P}_{\pm}}{\ud s \ud^{2}r^{\LCperp}}=\frac{\alpha}{2\pi \eta_p} \int \ud\vphi \left[\left(2yf_{s}-z\right)\textrm{Ai}(y)\right.\nonumber\\
  &\left.~~~~~~~\pm 2 f_{s}|\xi^{\LCperp}(\vphi)|^{-1}\left| \Pi^{\LCperp}(\vphi)\times\xi^{\LCperp}(\vphi)\right| \sqrt{z}\textrm{Ai}'(y) \right],
\label{Eq_Cir_LCFA}
\end{align}
in which $\xi^{\LCperp}(\vphi):=a'^{\LCperp}(\vphi)/m$ includes the field amplitude, direction and pulse envelope. In the same vein as the linearly-polarised case, we can integrate over the transverse photon momenta to acquire the lightfront momentum spectrum of the LCFA:
\begin{align}
\frac{\ud\tsf{P}_{\pm}}{\ud s}&=-\frac{\alpha}{2\eta_p}\int \ud\vphi~\left[\textrm{Ai}_{1}(z)+2f_{s}\frac{1}{z}\textrm{Ai}'(z)\right]\,,
\label{Eq_Cir_LCFA_s}
\end{align}
For a constant crossed field background, when one integrates over the transverse photon momentum, one is simultaneously integrating over the trajectory of the electron since there is a one-to-one mapping between the component of photon momentum parallel to the background field and the electron's phase position \cite{ritus85}. Due to the symmetry of a circularly-polarised background, after this integration over the trajectory, no information is retained about the polarisation of the background field. Therefore the spectrum in \eqnref{Eq_Cir_LCFA_s} is the same for photons in different circularly-polarised states (this is not the case for the angularly-resolved LCFA, for the same reason). The probability change due to scattering into different polarisation states: $\Delta \tsf{P}=\tsf{P}_{+}-\tsf{P}_{-}=0$. However, we note that when the intensity of the linearly polarised laser field is twice as large as the intensity of the circularly polarised laser field, the total unpolarised probability is indeed the same for linear and circular cases ($\tsf{P}=\tsf{P}_{1}+\tsf{P}_{2}=\tsf{P}_{+}+\tsf{P}_{-}$).

\subsection{Circularly polarised monochromatic field}
The derivation is very similar to the linearly-polarised monochromatic case, [where now $g(\phi) = 1$ in \eqnref{eqn:clinear1}] with the added simplification that $a^{2} = -m^{2}\xi^{2}$ is a constant and so only one Jacobi-Anger expansion is required for each phase integration, resulting in an expression with products of two Bessel functions (rather than products of four Bessel functions as in the linear case).
We find the probability $\tsf{P}_{\pm} = \sum_{n=-\infty}^{\infty} \tsf{P}_{\pm,n}$, where
 \begin{align}
\tsf{P}_{\pm,n}&=\frac{\alpha N_{\phi}}{2\eta^2_p}\int^{1}_{0} \ud s \int \ud^{2}r^{\LCperp}\frac{s}{t}\delta\left(c+4\beta-n\right) \nonumber\\
&\left\{f_{s}\xi^2/2\left(J^2_{n+1}+J^{2}_{n-1}-2J^2_n\right) -J^{2}_n\right. \nonumber\\
   &\left.\pm2f_{s}\xi^2\left[s(1 +\xi^2)/(\eta_p t)-n\right]J'_{n}J_n/\zeta_{c}\right\}\,,
\end{align}
and the argument of the Bessel functions $J_{n}\equiv J_{n}(\zeta_{c})$ $[J'_n\equiv J'_n(\zeta_{c})]$ is
\[\zeta_{c}= \xi\, |\ell^{\LCperp}|\frac{s}{\eta_{p} t}.\]
The delta function again fixes the net number of absorbed laser photons $n>0$ and gives the kinematic range of $n$th harmonic, $0<s<s_{c,n}$, where:
\bea
s_{c,n}=\frac{2n\eta_p}{2n\eta_{p}+1+\xi^2}\,,
\label{Eq_harmonic_region_cir}
\eea
which is the linear case \eqnref{Eq_harmonic_region_lin} with the substitution $\xi^2/2 \to \xi^2$.
\subsubsection{Circularly-polarised angular spectrum}
Evaluating the delta function by integrating over $s$, we obtain the angular spectrum:
\begin{align}
&\frac{\ud^2\tsf{P}_{\pm,n}}{\ud r_x \ud r_y}=\frac{\alpha N_{\phi}}{2\eta^2_p}~\frac{\mathcal{C}^2_n}{(1+\mathcal{C}_n)^2}\frac{1}{n} \nonumber\\
&~~~~\left\{f_{s}\xi^2/2\left(J^2_{n+1}+J^{2}_{n-1}-2J^2_{n}\right)-J^{2}_n\right.\nonumber\\
&\left.~~~~\pm2f_{s}\xi^2\left[\mathcal{C}_n(1 +\xi^2)/\eta_p-n\right]J'_{n}J_n/\zeta_{c,n}\right\}\,,
\label{Eq_cir_harmonic_angular}
\end{align}
where
\[\mathcal{C}_n=\frac{2n\,\eta_p }{(\ell^{\LCperp})^2+1+\xi^2}\,,\]
and the argument of the Bessel functions becomes:
\[\zeta_{c}\to \zeta_{c,n} = \xi\, |\ell^{\LCperp}|~\mathcal{C}_n/\eta_{p}.\]

\subsubsection{Circularly-polarised lightfront momentum spectrum}
Integrating the delta function over the transverse momentum $r^{\LCperp}$, we acquire the energy spectrum:
 \begin{align}
\frac{\ud\tsf{P}_{\pm,n}}{\ud s}&=\frac{\pi\alpha N_{\phi}}{\eta_p}\left\{\frac{f_{s}}{2}\xi^2\left(J^2_{n+1}+J^{2}_{n-1}-2J^2_{n}\right) -J^{2}_n\right. \nonumber\\
&\left.\pm2f_{s}\xi^2\left(\frac{s}{t}\frac{1 +\xi^2}{\eta_p}-n\right)\frac{1}{\zeta_{c,n}}J'_{n}J_n\right\}\,,
\label{Eq_cir_harmonic_S}
\end{align}
where the argument of the Bessel functions is replaced:
\[\zeta_{c} \to \zeta_{c,n} = \xi\, |\ell^{\LCperp}_n|\frac{s}{\eta_{p} t}\,,~~~~~~~|\ell^{\LCperp}_n|=\sqrt{2n\eta_p\frac{s_{c,n} - s }{s\,s_{c,n}}}.\]
Summing over polarisation states, we recover the unpolarised formula \cite{landau4}.

\section{Numerical calculations}~\label{Sec_numerical}
In this section, we consider an example scenario of a head-on collision between an $8\,\trm{GeV}$ electron and an $8$-cycle (full width at half maximum, $11\,\trm{fs}$) laser pulse with intermediate intensity $\xi=1$ and frequency $\omega_0=1.55\,\trm{eV}$. The corresponding pulse envelope is $g(\phi)=\cos^2(\phi/4\sigma)$, with $|\phi|<2\pi \sigma$ and $\sigma=4$, and the electron energy parameter is $\eta_p=0.095$. This choice of parameters is motivated by upcoming high-energy experiments such as LUXE and E320.
In this parameter region, the scattered photons are collimated in the electron incident direction with a very small angular spread.
We refer to $\epsilon_1$ and $\epsilon_{+}$ as ``$E$-polarisation'' states, because they are almost parallel to background fields with polarisation $\eps_1$ and $\eps_+$ in the linear and circular cases respectively (almost parallel, because photons are emitted with finite (small) angles). $\epsilon_2$ and $\epsilon_{-}$ are then referred to as ``$B$-polarisation'' states, as they are (almost) parallel to the magnetic field in each case. The polarisation purity is defined as the fraction of the $E$-polarised photons: $\mathcal{P}=\tsf{P}_{1}/\tsf{P}$ in the linear case and $\mathcal{P}=\tsf{P}_{+}/\tsf{P}$ in the circular case. For the $n$th order of harmonic, the polarisation purity is $\mathcal{P}_{n}=\tsf{P}_{1,n}/(\tsf{P}_{1,n}+\tsf{P}_{2,n})$ in the linear case and $\mathcal{P}_{n}=\tsf{P}_{+,n}/(\tsf{P}_{+,n}+\tsf{P}_{-,n})$ in the circular case.

\subsection{Linearly polarised background}
\begin{figure}[t!!]
 \includegraphics[width=0.49\textwidth]{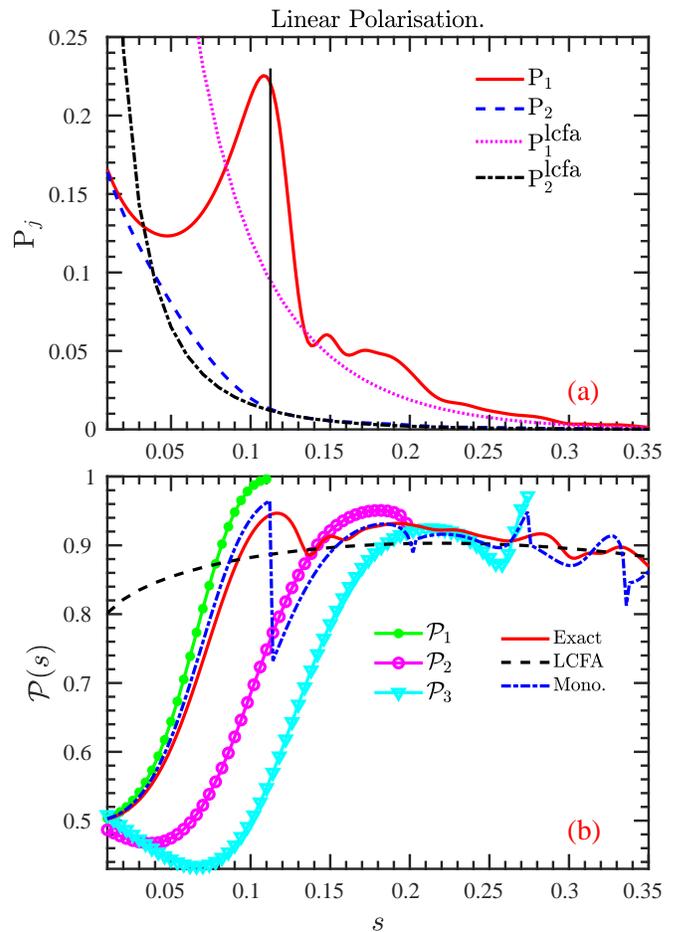}
\caption{(a) Polarised energy spectra: exact QED results and LCFA results. The black vertical solid line denotes the upper limit of the first harmonic $s_{\sscript{l},1}$ from Eq.~(\ref{Eq_harmonic_region_lin}).
(b) Polarisation purity: exact QED result, LCFA result and monochromatic result. The markers denotes the polarisation purity of the first three harmonics $\mathcal{P}_n$. Head-on collision is considered with the parameters: $\xi=1$, $\omega_0=1.55~\textrm{eV}$.}
\label{Fig_lin_polar_prob_purity}
\end{figure}
Fig.~\ref{Fig_lin_polar_prob_purity} shows the energy spectra of the polarised photons and the behaviour of the polarisation purity for different value of $s$ in the linear background field.
As shown in (a), the relative values of the polarised spectra depend sensitively on the photon lightfront momentum $s$:
in the low-energy limit $s\to 0$, the photons are unpolarised ($\mathcal{P}=0.5$) as shown in (b). However, in the high-energy region, starting around the Compton edge [vertical black line in Fig.~\ref{Fig_lin_polar_prob_purity} (a), the kinematic bound of the first harmonic $s_{\sscript{l},1}$], photons are more likely to be scattered into the $E$-polarisation state, reaching a high purity $>90\%$.

An evident harmonic structure can be observed in the photons' energy spectra in Fig.~\ref{Fig_lin_polar_prob_purity} (a) for the $8$-cycle laser pulse. The harmonic structure corresponds to the multi-peak structure in the polarisation purity [red solid line in Fig.~\ref{Fig_lin_polar_prob_purity} (b)]. To illustrate this, we plot the polarisation purity (blue dash-dotted line) from a monochromatic background, which broadly agrees with the pulsed result. We can also see that the highest purity (about $95\%$) results from the first harmonic, which is dominant in the $E$-polarisation state and becomes purely $E$-polarised at the Compton edge. For higher order harmonics, the polarisation purity becomes smaller, and in the low lightfront momentum region can even display a dominance of $B$-polarised photons where $\mathcal{P}<0.5$, for $\mathcal{P}_{2,3}$ in Fig.~\ref{Fig_lin_polar_prob_purity} (b).

\begin{figure}[t!!]
\center{\includegraphics[width=0.49\textwidth]{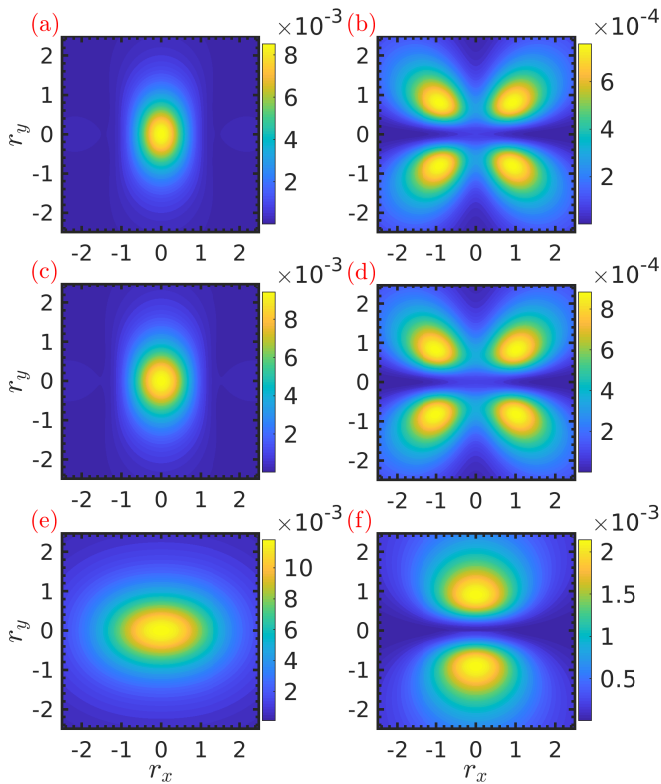}}
\caption{Angular distribution $d^2\tsf{P}/dr_xdr_y$ of the linearly polarised photons. Left column: $E$-polarisation; Right column: $B$-polarisation. Upper panel: (a) and (b), exact QED calculation for a pulse. Central panel: (c) and (d), QED calculation for a monochromatic field.  We take $N_{\phi}=4$; Bottom panel: (e) and (f), the LCFA results. Same parameters are used in Fig.~\ref{Fig_lin_polar_prob_purity}.}
\label{Fig_lin_polar_Angular_Exact}
\end{figure}

\begin{figure}[t!!]
\center{\includegraphics[width=0.49\textwidth]{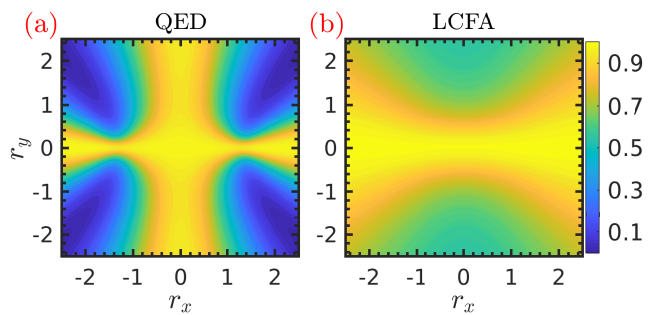}}
\caption{Angular distribution of the polarisation purity $\mathcal{P}$ in linearly polarised background. (a) the exact QED calculation for a pulse; (b) the LCFA results. Same parameters are used in Fig.~\ref{Fig_lin_polar_prob_purity}.}
\label{Fig_Lin_purity}
\end{figure}

In Fig.~\ref{Fig_lin_polar_Angular_Exact}, we show the angular spectra of the polarised photons, $d^2\tsf{P}/dr_xdr_y$ for the same parameters as in Fig.~\ref{Fig_lin_polar_prob_purity}. [Only photons with $0<s<0.4$ have been included as higher values of $s$ are strongly suppressed for these parameters, as shown in Fig.~\ref{Fig_lin_polar_prob_purity} (a)].  We observe that $E$-polarised photons are tightly collimated with the incident electron propagation direction $(r^{\LCperp}=0)$ with an angular spread: $2|r^{\LCperp}|/\gamma_p\sim 2\xi/\gamma_p\approx 0.13\,\trm{mrad}$. This is much narrower than the $B$-polarised photons' distribution, and with the peak value one order of magnitude larger than the $B$-polarised photons.
We also see that the $B$-polarised photons are scattered in a four-peak quadrupole distribution, while the $E$-polarised photons are scattered into the typical dipole-radiation distribution. In order to compare with the monochromatic results, we pick a scaling factor $N_{\phi}$ by replacing the integration over the phase with an integral over the pulse envelope, $N_{\phi} = \int d\phi/2\pi \to \int d\phi \,g(\phi) /2\pi = 4$. In Figs. (c) and (d), we see good quantitative agreement with the pulsed results and also that the multipole-structure is clearly reproduced.

The significant difference in the angular distribution of the polarised photons results in the particular structure in the distribution of the photon polarisation purity shown in Fig.~\ref{Fig_Lin_purity} (a). Around the electron incident direction: $|r^{\LCperp}|<0.5$ corresponding to the photon angular spread $\sim0.064\,\trm{mrad}$, we can achieve a almost purely $E$-polarised ($\mathcal{P}> 96\%$) photon beam, and for a broader angular spread $\sim0.1\,\trm{mrad}$, photons are emitted with the polarisation purity of $85\%$. Here we want to emphasise that because high energy photons are tightly collimated in the electron incident direction, a highly $E$-polarised $\gamma$-ray can be generated if we exclude the photons with a large scattering angle \cite{king20a}.

For comparison, we also show the corresponding LCFA results in Figs.~\ref{Fig_lin_polar_prob_purity},~\ref{Fig_lin_polar_Angular_Exact} and~\ref{Fig_Lin_purity}. As expected~\cite{king19a}, in this intermediate energy region, LCFA cannot reproduce the harmonic-structure in the spectra, and therefore cannot be used in this regime to give an accurate prediction of the purity of the emitted photons. As shown in Fig.~\ref{Fig_lin_polar_prob_purity}, for lower-energy photons, $s\to 0$, the LCFA prediction for $d\tsf{P}/ds$ diverges, as is well known, and hence it overestimates the polarisation purity. For higher energy photons, the LCFA result averages through the harmonic structure. We can also see that the angularly-resolved LCFA result shows large deviations from the exact QED calculations shown in Fig.~\ref{Fig_lin_polar_Angular_Exact}. The LCFA broadens the angular distribution of $E$-polarised photons and merges the four-peaks structure in the distribution of the $B$-polarised photons into a double-peaks structure. Therefore, the LCFA predicts a significantly different polarisation purity distribution in Fig.~\ref{Fig_Lin_purity}. As the LCFA result overestimates the peak value of the photon distribution, especially for the $B$-polarised photons, the value of the polarisation purity in the $y$ direction (perpendicular to the field polarisation direction) is underestimated.

\subsection{Circularly polarised background}

\begin{figure}[t!!]
 \includegraphics[width=0.45\textwidth]{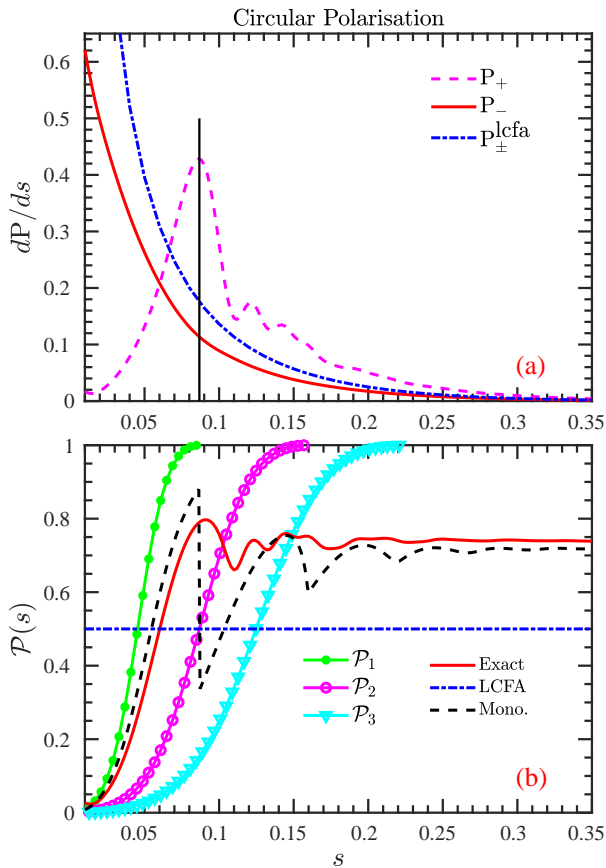}
\caption{(a) Polarised energy spectra: exact QED results and LCFA results. The black vertical solid line denotes the upper limit of the first harmonic $s_1$ from Eq.~(\ref{Eq_harmonic_region_cir}).
(b) Polarisation purity: exact QED result, LCFA result and monochromatic result. The markers denote the polarisation purity of the first three harmonics $\mathcal{P}_n$. Same parameter as in Fig.~\ref{Fig_lin_polar_prob_purity}.}
\label{Fig_Cir_spectrum}
\end{figure}
Fig.~\ref{Fig_Cir_spectrum} (a) shows the energy spectra of the circularly polarised photons in a circularly-polarised background field.
Unlike in a linearly-polarised background, the lower-energy photons, $s\to 0$, are purely $B$-polarised, $\mathcal{P}\to 0$, whereas higher-energy photons, with $s\gtrsim s_{c,1}$ [$s_{c,1}$ is the Compton edge, vertical black line in Fig.~\ref{Fig_Cir_spectrum} (a) ], are highly polarised in $E$-polarisation state.
Thus the polarisation purity increases from $\mathcal{P}=0$ at the low-energy limit up to $\mathcal{P}>70\%$ in the higher-energy region as shown in Fig.~\ref{Fig_Cir_spectrum} (b).
This tendency can be explained via the monochromatic result, which as shown in Fig~\ref{Fig_Cir_spectrum} (b) (black dash-dotted line), which matches well with the pulse result (red solid line). This change in the polarisation purity as $s$ is increased from $0$ can be explained mathematically. From Eq.~(\ref{Eq_cir_harmonic_S}), the perturbative expansion relation for small argument of the Bessel function, $J_n(\zeta)\sim \zeta^{n}$ can be used to show that in the $s\to 0$ limit for the $n$th order of harmonic gives  $\ud\tsf{P}_{\pm,n}/\ud s \sim 2f_{s}\xi_0(n^2\mp n^2)\zeta^{2n-2}_{c,n}$ whereas in the $s\to s_{c,n}$ limit, $\ud\tsf{P}_{\pm,n}/\ud s \sim 2f_{s}\xi_0(n^2\pm n^2)\zeta^{2n-2}_{c,n}$. Thus the harmonics are purely $B$-polarised ($\mathcal{P}_n=0$) at $s\to 0$ and purely $E$-polarised ($\mathcal{P}_n=1$) at the harmonic bound $s\to s_{c,n}$, as also shown in Fig~\ref{Fig_Cir_spectrum} (b).

\begin{figure}[t!!]
 \includegraphics[width=0.47\textwidth]{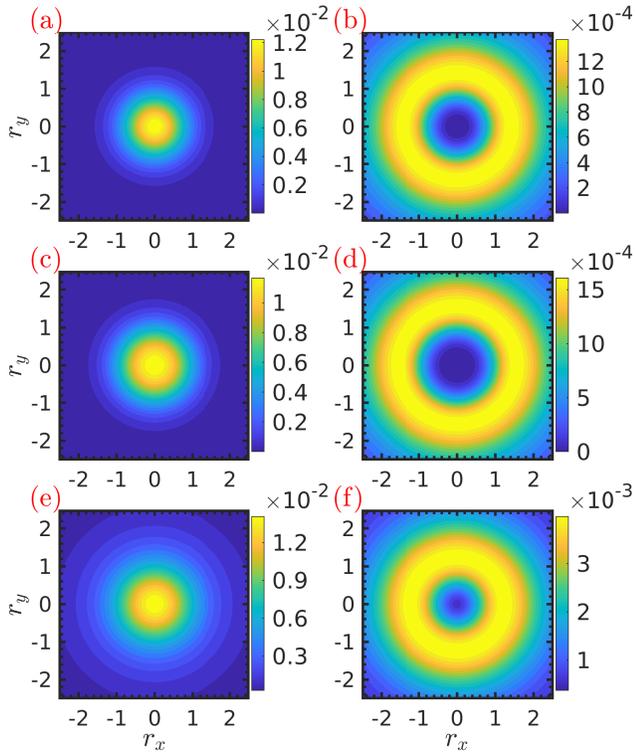}
\caption{Angular distribution $d^2\tsf{P}/dr_xdr_y$ of the circularly polarised photons. Left column: $E$-polarisation; Right column: $B$-polarisation. Upper panel: (a) and (b), exact calculation for a pulse; Central panel: (c) and (d), calculation for a monochromatic field ($N_{\phi}=4$); Bottom panel: (e) and (f), the LCFA results. Photons are included in the range $0<s<0.4$, and the parameters are the same as in Fig.~\ref{Fig_Cir_spectrum}.}
\label{Fig_Cir_Dist}
\end{figure}

\begin{figure}[t!!]
\center{\includegraphics[width=0.49\textwidth]{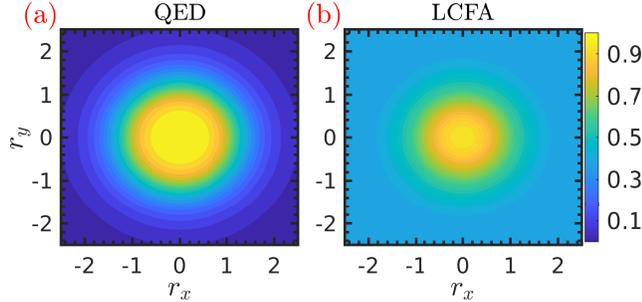}}
\caption{Angular distribution of the polarisation purity in the circularly polarised background. (a) the exact QED calculation for a pulse; (b) the LCFA results. Same parameters are used in Fig.~\ref{Fig_Cir_spectrum}.}
\label{Fig_Cir_purity}
\end{figure}

In Fig.~\ref{Fig_Cir_Dist}, we present the angular distribution of the polarised photons. As shown, the azimuthal symmetry is maintained in an $8$-cycle pulse. Similar to the linear case, the $E$-polarised photons are more collimated in the electron incident direction with an angular spread $2|r^{\LCperp}|/\gamma_p\sim 2\xi/\gamma_p\approx 0.13\,\trm{mrad}$, and the distribution peak value is one order of magnitude larger than the $B$-polarised photons. In \figref{Fig_Cir_purity}, we plot how the polarisation purity depends on scattering angle. We observe that the photons scattered with the smallest angle have the highest purity. For the current parameters, within an angular spread $<0.077\,\trm{mrad}$  the scattered photons are almost purely $E$-polarised with $\mathcal{P}> 96\%$  and within an angular spread of $\sim 0.1\,\trm{mrad}$ have a purity of $\mathcal{P}\sim 88\%$. Alternatively, we could concentrate on large-angle scattering and find that for angles $>0.33\,\trm{mrad}$, photons are highly $B$-polarised ($\mathcal{P}<5\%$) instead.

In Figs.~\ref{Fig_Cir_spectrum},~\ref{Fig_Cir_Dist} and~\ref{Fig_Cir_purity}, we also show the LCFA results for comparison. Away from the centre of the distribution in Fig.~\ref{Fig_Cir_purity}, the LCFA predicts a polarisation purity of $0.5$, which is equal to the ratio one acquires after integrating out transverse photon momenta, and shows that the LCFA cannot resolve angularly-resolve polarisations for wide-angle scattering. In the centre of the distribution, we see that the LCFA predits a narrower peak than the QED result, suggesting that the actual situation is more favourable in this regime than simulations based on the LCFA would predict. Because the LCFA result overestimates the value of the $B$-polarised photon distribution, the peak value of the polarisation purity in Figs.~\ref{Fig_Cir_purity} is underestimated to be less than the exact QED value: $\mathcal{P}<96\%$.

\section{Classical analysis}~\label{Sec_classical}
\begin{figure}[h!!]
\includegraphics[width=0.48\textwidth]{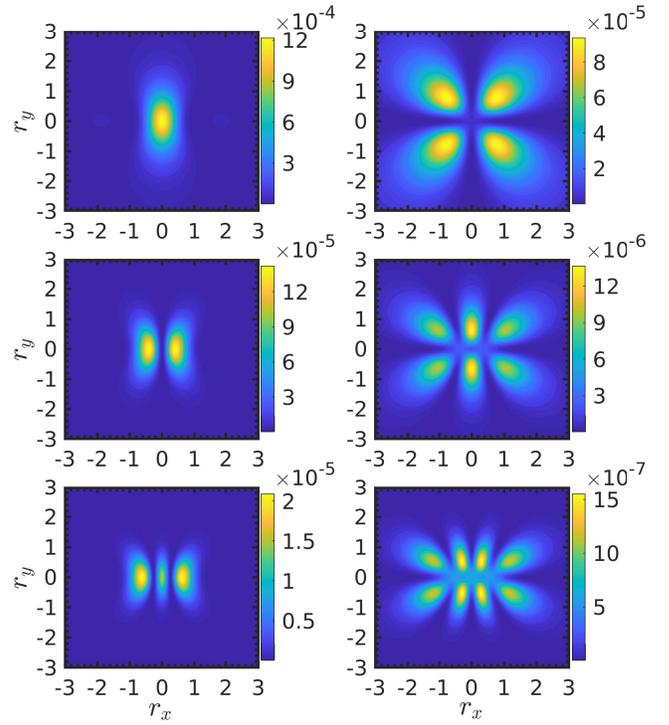}\\
\caption{Angular spectra of the first three harmonics in the linearly polarised monochromatic field: $\xi=1$, $\eta_{p}=0.095$ ($8\,\trm{GeV}$, electrons) for photons polarised in state $\epsilon_{1}$ (parallel to the field, left column) and $\epsilon_{2}$ (perpendicular to the field, right column). Upper panel: first harmonic; Central panel: second harmonic; Bottom panel: third harmonic.}
\label{Harmonic_Lin_distribution}
\end{figure}

In Fig.~\ref{Fig_lin_polar_Angular_Exact} (a) and (d), we observe the multipole-radiation structure in the angular spectra of the polarised photons. To show this structure more clearly, we present the angular distribution of the first three harmonics in Fig.~\ref{Harmonic_Lin_distribution}. As shown, the angular distributions of the different harmonics display different orders of multipole-radiation \cite{jackson99}.

In order to interpret the origin of the orbital-type shapes in the angular spectra, which persist when the quantum parameter $\chi_{p} \ll 1$,  we perform a classical calculation of nonlinear Thomson scattering. If the classical analogue of the  differential probability of photon emission $d^{3}\tsf{P}^{\tsf{cl.}}/dk^{3}$ is given by the energy of the emitted field in units of frequency:
\[
\frac{d^{3}\tsf{P}^{\tsf{cl.}}}{dk^{3}} = \frac{1}{k^{0}}\frac{d^{3}P^{0}}{dk^{3}},
\]
where $P^{\mu}$ is the four-momentum of the field emitted by the electron through interaction with the electromagnetic background. Proceeding from Coleman \cite{coleman82}, we write:
\bea
 \tsf{P}^{\tsf{cl.}}  = \frac{2}{(2\pi)^{3}}\int d^{4}k~ \widetilde{\j}^{\ast}_{\lambda}(k) \widetilde{\j}^{\lambda}(k) \theta(k^{0})\delta(k^{2}),
\eea
where $\widetilde{\j} = \int d^{4}x~ j(x)\exp[ik\cdot x]$ and we use a definition for the classical current of \mbox{$j^{\mu}(x) = e \int \delta^{(4)}(x-x'(s)) \pi'^{\mu}(\tau)\,d\tau$}, where $\tau$ is the proper time, related to the external-field phase by $d\tau/d\vphi = (m\eta_{p})^{-1}$. This then leads to:
\bea
\tsf{P}^{\tsf{cl.}}  = \frac{\alpha}{\pi^{2}\eta_{p}^{2}}\int \frac{ds\,d^{2}k^{\LCperp}}{2\,s}
 \,d\vphi\,d\vphi'~\,\pi(\vphi)\cdot\pi(\vphi')
\,\mbox{e}^{i\,k\cdot[x(\vphi)-x(\vphi')]}. \nn \\
\eea
At this point, we decompose the trajectory using the basis introduced in \eqnref{eqn:polbasis} and the derivation is then very similar to the QED version. The total probability can be written $\tsf{P}^{\tsf{cl.}} = \tsf{P}^{\tsf{cl.}}_{1} + \tsf{P}^{\tsf{cl.}}_{2}$, where $\tsf{P}^{\tsf{cl.}}_{j} = \sum_{n}\tsf{P}^{\tsf{cl.}}_{j,n}$ and:
\bea
\tsf{P}^{\tsf{cl.}}_{j} &=& \frac{2\alpha}{\eta_{p}}\int \frac{d\vphi}{2\pi}\int s\,ds\,d^{2}r^{\LCperp}\delta(\tilde{c}+2\tilde{\beta}_{s}-n) \nn \\
&&
\left.\times \left\{\delta_{j,2}(r\cdot\eps_{2})^{2}\Gamma_{0,n}^{2}+\right. \right. \nn \\
&& \left. \delta_{j,1}\left[(r\cdot \eps_{1})^{2}\Gamma_{0,n}^{2}+2m\xi\,r\cdot\eps_{1}\,\Gamma_{1,n}\Gamma_{0,n} + \xi^{2}\Gamma_{1,n}^{2}\right]\right\}, \nn \\\label{eqn:class2}
\eea
where the arguments of the $\Gamma_{l,n}$ functions are now:
\[
\tilde{\alpha}_{s} = \frac{\xi\,s}{\eta_{p}}\,r\cdot\eps; \qquad \tilde{\beta}_{s} = \frac{\xi^{2}s}{8\eta_{p}}.
\]
We note that energy-momentum conservation is different in the classical case, because the electron experiences no recoil classically. This is reflected by the modified delta-function argument:
\[
 \delta(\tilde{c}+2\tilde{\beta}_{s}-n) = \delta\left[\frac{s}{2\eta_{p}}\left(1+\frac{\xi^{2}}{2} + (\mbf{r}^{\LCperp})^{2}\right) -n \right],
\]
which on comparison with the quantum version \eqnref{eqn:delt1} has been modified by a replacement $s/(1-s) \to s$. (Recalling that the scattered electron energy parameter $\eta_{q} = \eta_{p}(1-s)$, we see that setting $1-s \to 1$, is equivalent to neglecting the electron recoil.) There is also an extra term in the classical integrand which originates from the  $\mbox{e}_{\pm}$ directions, of the form:
\[
 [1+(\mbf{r}^{\LCperp})^{2}]\Gamma_{0,n}^{2} - 2\xi\, r\cdot\eps_{1} \Gamma_{0,n}\Gamma_{1,n} + \xi^{2}\Gamma_{0,n}\Gamma_{2,n},
\]
but this disappears, because it is a boundary term in the integration over $\phi$ and $\phi'$ (it is proportional to $k\cdot \pi$, which occurs in the exponent).
\newline

One immediate consequence arises in the angular spectrum when $s$ is integrated over in \eqnref{eqn:class2}.
\bea
\tsf{P}^{\tsf{cl.}}_{j} &=& \frac{2\alpha}{\eta_{p}}\int \frac{d\vphi}{2\pi}\int_{0}^{2\pi}\frac{d\theta_{r}}{2\pi}\int_{r_{n}}^{\infty} dr\,r \,s_{n}^{\ast}\left\{\delta_{j,2}(\mbf{r}^{\LCperp}\cdot\pmb{\eps}_{2})^{2}\Gamma_{0,n}^{2}+\right. \nn \\
\label{eqn:Pcl1}
&&
\left.\!\!\delta_{j,1}\left[(\mbf{r}^{\LCperp}\cdot \pmb{\eps}_{1})^{2}\Gamma_{0,n}^{2}-2m\xi\,\mbf{r}^{\LCperp}\cdot\pmb{\eps}_{1}\,\Gamma_{1,n}\Gamma_{0,n} + \xi^{2}\Gamma_{1,n}^{2}\right]\right\} \nn \\\label{eqn:class2}
\eea
where $s_{n}^{\ast} = y_{n}^{\ast}$ from \eqnref{eqn:yast1}. However, because there is no electron recoil in the classical description, \mbox{$s_{n}^{\ast} \leq 1$}, whereas the equivalent QED parameter, $y_{n}^{\ast}$, is unbounded. This means that there exists a classical ``angular edge'' in analogy with the Compton edge \cite{harvey09} but in the angular spectrum:
\bea
r_{n} = \theta(u_{n})\sqrt{u_{n}}; \qquad u_{n} = 2n\eta_{p}-1-\xi^{2}/2.
\eea
Therefore, the smoothing of the angular harmonic edges is a quantum effect.

Upon comparison with the quantum version for the total probability \eqnref{eqn:quant1}, we note two further differences. First, the decomposition that we have made to the electron's trajectory, by projecting the instantaneous \emph{electron momentum} onto the directions $\eps_{1,2}$ matches the decomposition in the quantum case, $\epsilon_{1,2}$ of the \emph{photon polarisation}. This mainly originates from the choice of polarisation basis, in which $p\cdot\epsilon_{1} = r\cdot \eps_{1}$ and $p\cdot\epsilon_{2}=r\cdot\eps_{2}$. We see thereby, the mapping between the classical electron's motion and the polarisation of the emitted photon in the QED case. The second difference, is that there are terms from QED that are absent in the classical formula, in particular the term:
\[
 -2(m\xi)^{2}\frac{s^{2}}{4(1-s)}\,\left[\Gamma_{2,n}\Gamma_{0,n} -\Gamma_{1,n}^{2}\right].
\]
This term is of purely quantum origin, which is made manifest when we recall that $s= \eta_{k}/\eta_{p} = \vkap\cdot k/\vkap\cdot p \propto \hbar$, and so when we take the limit of $\hbar \to 0$ of the quantum expression, this term disappears. (Incidentally, the other terms survive, because we have the combination $\eta_{p}^{-1}\int ds$, which survives in the limit of $\hbar \to 0$.)
\newline

The angular harmonic spectrum of photons emitted with polarisation perpendicular to the field resembles the spherical harmonic decomposition of the Green's function of the wave equation from classical electrodynamics~\cite{jackson99}. A connection to this expansion can be made by noting that the leading term dictating the shape of the angular harmonics is given by the $\Gamma_{0,n}^{2}$ where:
\bea
\Gamma_{0,n}=\sum_{t=-\infty}^{\infty}J_{t}\left(\frac{\xi^{2}s}{8\eta_{p}}\right)J_{2t+n}\left(\frac{\xi\,s\,r~\cos\theta}{\eta_{p}}\right).
\eea
For the lowest harmonics $s\ll 1$ and our analysis leads to the highest polarisation purities when $\xi/\eta_{p}$ is not too large, so the Bessel arguments are, in general, small. It turns out that the limit \cite{watson22}:
\bea
J_{n}(z) = \lim_{m\to\infty} m^{n}P_{m}^{-n}\left(\cos\frac{z}{m}\right) ,
\eea
where $P_{m}^{-n}$ are the associated Legendre polynomials,  is already well-approximated for small argument $|z|\ll 1$ at $m=1$. Then we see that the leading term can be approximated by:
\bea
\Gamma_{0,n}=\sum_{t=-\infty}^{\infty}J_{t}\left(\frac{\xi^{2}s}{8\eta_{p}}\right)P^{-(2t+n)}_{1}\left[1-\left(\frac{\xi\,s\,r~\cos\theta}{\eta_{p}}\right)^{2}\right]. \nn \\
\eea
Upon comparison with the form of the spherical harmonics:
\bea
Y_{l}^{m}(\theta,\phi) = N\mbox{e}^{im\phi} P_{l}^{m}(\cos\theta),
\eea
and noting that $P_{l}^{-m} \propto P_{l}^{m}$, we see that each higher harmonic $n$ is associated with a higher value of the index $m$.

\section{Conclusion}~\label{Sec_conclusion}
  We investigated the polarisation of a photon generated by nonlinear Compton scattering of an electron in a plane wave background.
  Our analysis considers the generation of linearly-polarised and circularly-polarised photons in correspondingly polarised background fields.
  We considered a finite pulse, a monochromatic background and the locally constant field approximation (LCFA) for each polarisation case.
  The lightfront-momentum and angularly-resolved spectra for each case were presented.

  Motivated by upcoming high-energy experiments LUXE at DESY and E320 at FACET-II, we focused attention on having a plane-wave laser pulse of intermediate intensity parameter $\xi=1$. We found that the harmonic structure of the photon spectrum is reflected in the polarisation purity of the scattered photons.  The angular spectrum of emitted photons is substantially different for photons polarised parallel to the electric field ($E$-polarised) to the angular spectrum of photons polarised parallel to the magnetic field ($B$-polarised), in both linear and circularly-polarised cases. By performing a classical calculation for the equivalent process of nonlinear Thomson scattering, we identified an explanation for the different angular distribution of different photon polarisation states due to the motion of the electron in a plane-wave background in the electric and magnetic field directions.

  Approaches based on the LCFA (which is the main method by which quantum effects are included in numerical simulations of high-intensity laser-matter interactions), cannot reproduce the structure of the angular distributions. In particular for the linearly polarised case, the multipole structure and for the circularly-polarised case, the spherical harmonic structure, are both beyond local approaches. The LCFA approach is thus incapable of describing the angular distribution of the photon polarisation purity in the parameter regime of interest in high-energy  experiments at intermediate field intensity.
  The LCFA also performs poorly at predicting the yield of photons with a given polarisation in the low-energy part of the spectrum as well as around the first harmonic (the Compton edge), which is an identified experimental observable of interest \cite{Abramowicz:2019gvx}.
  Finally, we conclude that the significant difference in the angular and energy dependency of $E$- and $B$-polarised photons lends itself to the possibility of generating highly-brilliant, highly polarised sources of gamma photons, as explored in \cite{king20a}.

\acknowledgments
S. T. and B. K. are supported by the UK Engineering and Physical Sciences Research Council, Grant No. EP/S010319/1.
\appendix

\section{Regularisation of $\theta^{-2}$ integrals}\label{AppendixA}
In evaluating the total probability of polarised $1\to2$ processes like NLC or nonlinear Breit-Wheeler processes, we are faced with regularisation of the following integral:
\bea
 \mathcal{J} = \int_{-\infty}^{\infty} \frac{d\theta}{\theta^{2}} \e^{ig(\theta)},
\eea
where $g(-\theta) = -g(\theta)$.
We can write this in terms of trigonometric functions $\mathcal{J} = \mathcal{J}_{c} + i \mathcal{J}_{s}$ and then proceed to use the trick of introducing a slight contour rotation:
\[
 \mathcal{J} = \lim_{\eps\to0} \int_{-\infty}^{\infty} \frac{d\theta}{\theta + i \eps}\frac{1}{\theta+i\eps}~ \e^{ig(\theta)}.
\]
Then $\mathcal{J}_{s}$ can be dealt with in the following way:
\onecolumngrid
\bea
 \mathcal{J}_{s} &=& \lim_{\eps\to0} \int_{-\infty}^{\infty} \frac{d\theta}{(\theta + i \eps)^{2}}~ \sin\left[g(\theta)\right] \nn \\
  &=& \lim_{\eps\to0} \left\{\int_{-\infty}^{\infty} \frac{d\theta}{(\theta + i \eps)^{2}}~ \left(\sin\left[g(\theta)\right]-\theta\,g'(0)\right) + g'(0)\int_{-\infty}^{\infty} d\theta~\frac{\theta}{(\theta + i \eps)^{2}}\right\} \nn \\
  &=& \lim_{\eps\to0} \left\{\int_{-\infty}^{\infty} \frac{d\theta}{\theta^{2}}~ \left(\sin\left[g(\theta)\right]-\theta\,g'(0)\right) + g'(0)\int_{-\infty}^{\infty} d\theta~\left[\frac{-i\eps}{(\theta+i\eps)^{2}} + \frac{1}{\theta+ i \eps} \right] \right\} \nn \\
  &=& - i\pi g'(0),
\eea
where in the last line, we have used the fact that the first integral is identically zero (the Taylor series comprises entirely odd terms integrated over an even interval, and the Taylor series has an infinite radius of convergence for the  $\sin$ function).
\newline

The $\mathcal{J}_{c}$ integral can be performed by repeated use of the Sokhotski-Plemelj theorem:
\bea
 \mathcal{J}_{c} &=& \lim_{\eps\to0} \int_{-\infty}^{\infty} \frac{d\theta}{(\theta + i \eps)}\frac{1}{\theta+i\eps}~ \cos\left[g(\theta)\right] \nn\\
 &=& \lim_{\eps\to0} \int_{-\infty}^{\infty} d\theta \left[-i\pi \delta(\theta) + \mathcal{P} \frac{1}{\theta} \right] \frac{1}{\theta+i\eps}~ \cos\left[g(\theta)\right] \nn \\
  &=&  \lim_{\eps\to0}  \left\{- \frac{\pi}{\eps} + \int_{-\infty}^{\infty} d\theta   \frac{1}{\theta}  \frac{1}{\theta+i\eps}~ \left(\cos\left[g(\theta)\right]-1\right) + \mathcal{P}  \int_{-\infty}^{\infty} d\theta   \frac{1}{\theta}  \frac{1}{\theta+i\eps} \right\}\nn \\
    &=&  \lim_{\eps\to0}  \left\{- \frac{\pi}{\eps} + \int_{-\infty}^{\infty} d\theta   \frac{1}{\theta}  \frac{1}{\theta+i\eps}~ \left(\cos\left[g(\theta)\right]-1\right)  - \frac{i}{\eps} \mathcal{P}  \int_{-\infty}^{\infty} d\theta   \left[\frac{1}{\theta}  - \frac{1}{\theta+i\eps}\right] \right\}\nn \\
     &=&  \lim_{\eps\to0}  \left\{- \frac{\pi}{\eps} + \frac{\pi}{\eps} + \int_{-\infty}^{\infty} d\theta   \frac{1}{\theta}  \frac{1}{\theta+i\eps}~ \left(\cos\left[g(\theta)\right]-1\right)  \right\}\nn \\
     &=&   \int_{-\infty}^{\infty} d\theta   \frac{1}{\theta^{2}}\left(\cos\left[g(\theta)\right]-1\right).
    \eea
Then in total we have:
\bea
\mathcal{J} =  \pi g'(0) + \int_{-\infty}^{\infty} d\theta   \frac{1}{\theta^{2}}\left(\cos\left[g(\theta)\right]-1\right),
\eea
which reproduces the correct zero-field limit:
\bea
\lim_{\xi\to0}\mathcal{J} =  \pi g'(0) + \int_{-\infty}^{\infty} d\theta   \frac{1}{\theta^{2}}\left(\cos\left[g'(0) \theta\right]-1\right) = 0 .  \label{eqn:Jint}
\eea

\twocolumngrid

\bibliographystyle{apsrev}
\bibliography{current}

\end{document}